\newif\ifdouble
\newif\ifsingle
\newif\ifchange

\documentclass[sigconf, usenames, dvipsnames]{acmart}
\usepackage{listings}
\usepackage{xcolor} 

\usepackage{dblfloatfix} 
\usepackage{booktabs} 
\usepackage{arydshln}
\usepackage{subfig}
\newcommand{\remove}[1]{{\color{red}{\sout{#1}}}}

\renewcommand{\remove}[1]{}

\AtBeginDocument{%
  \providecommand\BibTeX{{%
    \normalfont B\kern-0.5em{\scshape i\kern-0.25em b}\kern-0.8em\TeX}}}

\author{Keiichi Ihara}
\affiliation{%
\institution{University of Colorado Boulder}
\city{Boulder}
\state{Colorado}
\country{United States}}
\email{keiichi.ihara@colorado.edu}

\author{Tianle Li}
\affiliation{%
\institution{University of Colorado Boulder}
\city{Boulder}
\state{Colorado}
\country{United States}}
\email{tianle.li@colorado.edu}

\author{Yasuhisa Shiino}
\affiliation{%
\institution{University of Colorado Boulder}
\city{Boulder}
\state{Colorado}
\country{United States}}
\email{yash3286@colorado.edu}

\author{Ryo Suzuki}
\affiliation{%
\institution{University of Colorado Boulder}
\city{Boulder}
\state{Colorado}
\country{United States}}
\email{ryo.suzuki@colorado.edu}

\begin{document}

\newcommand{\system}{MemoryDiorama}
\newcommand{\concept}{\textit{augmented memory cues}}
\newcommand{\Concept}{\textit{Augmented memory cues}}
\title{\system{}: 
Generating Dynamic 3D Diorama from
Everyday Photos for Memory Recall
}

\renewcommand{\shortauthors}{Ihara, et al.}
\begin{abstract}

We present \system{}, a prototype system that introduces \concept{}, a concept that extends captured personal media with AI-generated contextual information to enhance autobiographical memory recall.
\system{} transforms everyday photos into
dynamic 3D dioramas in mixed reality by integrating LLM-based scene analysis with 3D object generation, animation, and spatial composition. The system extracts geographic information, object attributes, lighting conditions, and atmospheric elements from the photos. 
It then animates these elements with generative components such as object animations, human motion, geographical effects, and particle effects to provide richer cues for memory recall. We evaluated \system{} in a within-subject user study with 18 participants, comparing three conditions: Photo-Only, Static Diorama, and \system{}. Compared with both Photo-Only and Static Diorama, \system{} elicited more internal and in-cue details during recall. It also increased perceptual details and visual vividness ratings, suggesting richer recollective experience.

\end{abstract}

\ccsdesc[500]{Human-centered computing~Mixed / augmented reality}

\keywords{Mixed Reality; Autobiographical Memory; Generative AI}

\begin{teaserfigure}
\centering
\includegraphics[width=\textwidth]{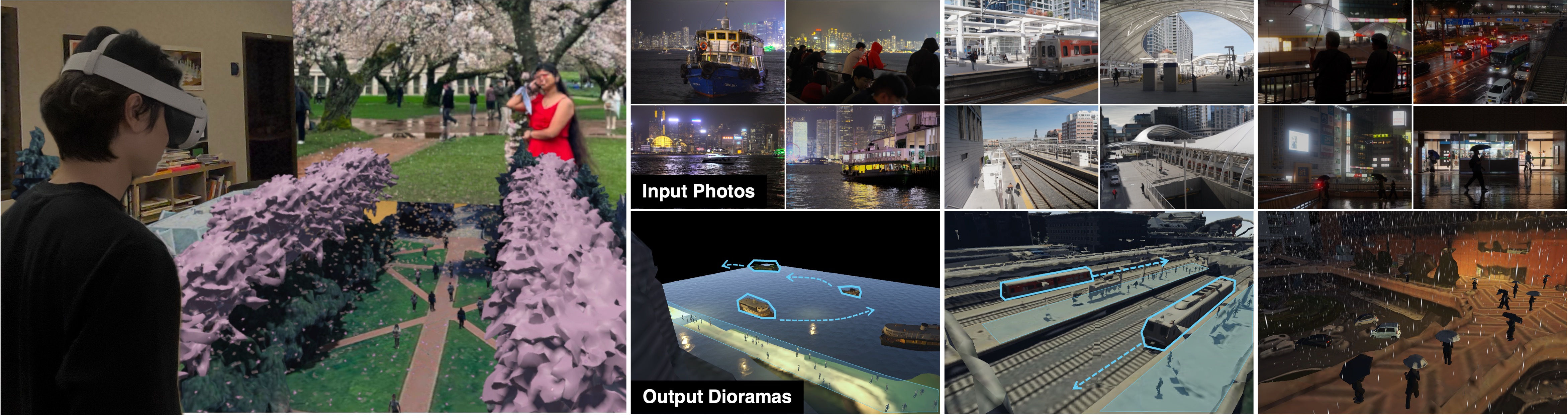}
\caption{
\system{} transforms everyday photos into dynamic 3D dioramas for autobiographical memory recall in mixed reality (left). Our system augments captured content with multiple generative cue layers, including object animations, human motion, particle effects, lighting effects, and geographical effects (right).
}
\label{fig:teaser}
\end{teaserfigure}

\maketitle

\section{Introduction}
Everyday photos are commonly used to revisit personal memories~~\cite{mcgookin2019reveal, chen2023photoclock, chen2023exploring}. But photos preserve only fragments of lived experience: a fixed view, a single moment, and a limited slice of the surrounding context. As a result, they often cue what was captured while providing little support for recalling what happened just outside the frame or in the broader atmosphere of the event.

To address these limitations, we introduce the concept of \concept{}, which we define as memory cues that extend captured personal media by generating additional contextual information with generative AI. Grounded in a user’s own photos, these cues retain a connection to the original lived event while providing a more diverse set of cues for recall. These cues can take multiple forms, including spatial expansion beyond the original frame, reconstructing moments as 3D scenes, and sensory augmentation through soundscape, tactile sensation, or smell. Rather than merely replaying recorded content, they computationally expand it into richer cues that may support recall of aspects of an experience that were not directly captured.

To demonstrate one form of \concept{}, we present \system{}, a system that transforms everyday photos into dynamic 3D dioramas. Each diorama is composed of five contextual layers: a geographical layer, an object layer, a human layer, a lighting layer, and a particle layer. Imagine recalling a quiet spring afternoon spent walking through a park (Figure~\ref{fig:teaser} (left)). The geographical layer captures the grassy ground and the gentle ripples of a nearby pond. The object layer brings in cherry blossom trees in full bloom, tinting the scene pink. As you walk beneath them, the particle layer fills the air with drifting petals. Looking around, the human layer places people taking photos and enjoying leisurely strolls. The lighting layer casts bright midday sunlight across the park, evoking the warmth of the day. Together, these layers transform a collection of photos into a richer and more dynamic reconstruction of a remembered event.

To realize this, we designed a three-phase pipeline consisting of (1) photo analysis, (2) element generation, and (3) placement and route generation. 
First, the system uses LLM-based scene analysis (Gemini) to analyze a collection of photos from a single autobiographical event and aggregate extracted semantic and spatial cues into a scene description, including scene context, visible objects and people, and environmental conditions. 
Next, the system generates the elements for each cue layer, including 3D object and human assets, environmental structures, and material textures, based on the scene descriptions. For object and human elements, it uses the image-to-3D model (SAM 3D) to convert extracted image regions into individual 3D elements, while generative image models (Nano Banana 2) are used to synthesize textures for geographical surfaces and particle-based effects. Finally, the system captures a top-down view of the 3D map and uses a multimodal LLM (Gemini) to annotate positions, areas, and routes for the generated elements. It then extracts these annotations with OpenCV, converts them into 3D spatial coordinates, and integrates the resulting scene into a dynamic diorama for mixed-reality exploration.

To examine how \system{} influences autobiographical memory recall, we conducted a within-subject user study with 18 participants comparing \system{} with two baseline conditions, Photo-Only and Static Diorama. Participants recalled personal events in each condition, and we evaluated both recalled details and subjective ratings. Results showed that \system{} elicited more internal and in-cue details than both baseline conditions, while also increasing perceptual details, visual vividness, and enjoyment.



Our main contributions are as follows:
\begin{enumerate}
  \item The concept of \concept{}, which uses generative AI to extend captured personal media into richer cues for autobiographical recall;
  \item \system{}, a system that transforms everyday photos into dynamic 3D dioramas;
  \item A user study showing that \system{} supports broader recollective elaboration during autobiographical recall.
\end{enumerate}
\section{Related Work}

\subsection{AR/VR Diorama Experiences}
Diorama representations have been widely explored in AR and VR to support interaction, spatial understanding, and experiential engagement with large-scale environments. 
Among these approaches, Worlds-in-Miniature (WiMs) have played an important role as an interaction technique that allows users to manipulate and explore virtual spaces through scaled-down representations~\cite{stoakley1995virtual, pausch1995navigation, danyluk2021design}. 

Originally introduced as an interaction technique for manipulating distant virtual objects, WiMs enable users to interact with large-scale environments through miniature representations (e.g., \textit{Virtual Reality on a WIM}~\cite{stoakley1995virtual}, \textit{Don't Panic}~\cite{bluff2019don}, and \textit{Poros}~\cite{pohl2021poros}). 
This approach has since been extended to interactions involving physical and hybrid spaces~\cite{stafford2006implementation, seo2016hybrid, tatzgern2015exploring, ihara2022virtual, ihara2022ar}.

WiMs have also been used as navigation aids, supporting indoor and outdoor~\cite{veas2012extended, mulloni2012indoor, niederauer2003non, kalkusch2002structured, hollerer2001user, fukatsu1998intuitive}, object-centric~\cite{pasewaldt2014multi, bonsch2016automatic, chittaro2005interactive}, virtual space~\cite{pausch1995navigation, laviola2001hands, bruder2009arch}, and temporal navigation~\cite{mahadevan2023tesseract}. 
Beyond navigation, WiMs have been applied to data exploration tasks, including visual comparison of spatial data (e.g., \textit{Worlds-in-Wedges}~\cite{nam2019worlds}), immersive data analysis (e.g., \textit{DataHop}~\cite{hayatpur2020datahop}), overview-and-detail exploration of volumetric data (e.g., \textit{Slice WIM}~\cite{coffey2011slice}), and correlation of geo-temporal datasets (e.g., \textit{GeoGate}~\cite{ssin2019geogate}).

Some work has explored dioramas for social and experiential purposes. 
For example, miniature avatars have been used to represent remote users for tutoring (e.g., \textit{Tutor In-sight}~\cite{thanyadit2023tutor}), remote communication (e.g., \textit{MiniMates}~\cite{kiuchi2025minimates}, \textit{3-D Live}~\cite{prince20023}, \textit{HoloBots}~\cite{ihara2023holobots}), and remote collaboration (e.g., \textit{Mini-Me}~\cite{piumsomboon2018mini}, \textit{On the Shoulder of the Giant}~\cite{piumsomboon2019shoulder}). 
Other work has explored tabletop dioramas for spectating experiences, such as watching sports (e.g., \textit{Soccer on Your Tabletop}~\cite{rematas2018soccer}) or racing events (e.g., \textit{Lapz’s F1 app}~\cite{LapzWatc30:online}).

Prior work on diorama representations suggests two key features for memory support: they can present diverse information within a compact, glanceable overview, and they allow users to inspect a scene from multiple angles through small bodily movements. Motivated by these properties, \system{} generates dioramas from personal photos to present diverse recall cues within a single scene and support revisiting the event from different perspectives.

\subsection{3D Scene Reconstruction and Generation}
Prior work on 3D scene representation spans faithful reconstruction, sparse-input 3D generation, and 3D world generation, varying in fidelity, input requirements, and grounding in lived experience.

3D reconstruction techniques have enabled spatial capture using multiple RGB-D cameras, as demonstrated in systems such as \textit{Holoportation}~\cite{orts2016holoportation}. Recently, neural scene representations such as Neural Radiance Fields (NeRF)~\cite{mildenhall2021nerf} and Gaussian Splatting~\cite{kerbl20233d} have significantly improved the fidelity of 3D reconstruction from RGB images.
In parallel, generative methods have emerged that produce
3D content from text or a single image rather than dense
multi-view captures. Text-to-3D approaches such
as \textit{DreamFusion}~\cite{poole2022dreamfusion} leverage
diffusion priors to generate 3D assets from text prompts,
while image-to-3D methods such as \textit{LRM}~\cite{hong2023lrm} reconstruct 3D geometry
from a single image. However, 3D reconstruction methods typically require dense visual input, while image-to-3D approaches remain limited in their ability to generate coherent scenes from sparse input.

At the other end of the spectrum, recent work explores generating entire 3D environments from text. \textit{Holodeck}~\cite{yang2024holodeck} uses an LLM to compose furnished 3D scenes by generating spatial constraints and retrieving objects from asset libraries. World models such as \textit{Genie3}~\cite{Genie3} and \textit{GWM-1}~\cite{Runway} generate interactive worlds. However, both directions generate environments from generic prompts and lack grounding in a user's lived experience, limiting their use for autobiographical memory recall.

In contrast to approaches that either reconstruct scenes as faithfully as possible or generate entirely fictional worlds, we take a compositional approach that is grounded in real personal experiences while leveraging generative AI to expand contextual cues.
Rather than attempting full scene recovery, our system reconstructs and augments multiple types of contextual elements to create dynamic 3D dioramas that balance autobiographical grounding with generative expressiveness.




\subsection{Systems for Personal Memory Recall}
People recall personal past events to reflect on their experiences, make sense of their lives, and connect with others~\cite{bluck2003autobiographical}. In response, interactive systems have supported personal memory recall in a variety of ways, from augmenting physical mementos to resurfacing digital traces and organizing personal archives.

Several approaches have focused on attaching digital memories to existing physical objects, such as memory boxes or souvenirs (e.g., \textit{The Memory Box}~\cite{frohlich2000memory}, \textit{Souvenirs}~\cite{nunes2008sharing}, \textit{Living Memory Box}~\cite{stevens2003getting}). These systems emphasize the role of tangible objects in supporting reflection on past experiences. 
\textit{TeleAbsence}~\cite{ishii2025teleabsence} further broadens this perspective by proposing memory-oriented experiences that evoke absent people and past moments through digital and physical traces, as illustrated by projects such as \textit{MirrorFugue}~\cite{xiao2010mirrorfugue}.

Van Gennip et al.~\cite{van2015things} found through a diary study that after physical objects, location was the most common cue that provoked reminiscing. Consistent with this, several research efforts have worked on creating reminiscence systems using location (e.g., \textit{Memory Tracer \& Memory Compass}~\cite{white2023memory}, \textit{Reveal}~\cite{mcgookin2019reveal}, \textit{Rewind}~\cite{tan2018rewind}). 
Other approaches have also been proposed to support reminiscence using different types of personal data, including social media (e.g., \textit{Pensieve}~\cite{peesapati2010pensieve}), email (e.g., \textit{MUSE}~\cite{hangal2011muse}), time (e.g., \textit{PhotoClock}~\cite{chen2023photoclock}, \textit{Chronoscope}~\cite{chen2023exploring}), audio (e.g., \textit{Slide2Remember}~\cite{kim2022slide2remember}, \textit{FM Radio}~\cite{petrelli2010fm}), and lifelogging data (e.g., \textit{SenseCam}~\cite{hodges2006sensecam, sellen2007life}).
Beyond individual reflection, some work has emphasized supporting reminiscence through interpersonal storytelling (e.g., \textit{Remini}~\cite{jiang2025remini}, \textit{MomentMeld}~\cite{kang2021momentmeld}, Li et al.'s work~\cite{li2023exploring}). 

Previous systems have been effective at resurfacing, organizing, and sharing personally captured traces for memory recall. However, they generally rely on cues that were explicitly recorded, rather than reconstructing additional contextual elements beyond the captured data. In contrast, our approach uses generative AI to expand such cues while remaining grounded in the original experience.


\section{Augmented Memory Cues}

\subsection{Definition and Scope}
\begin{figure}[h]
\centering
\includegraphics[width=\linewidth]{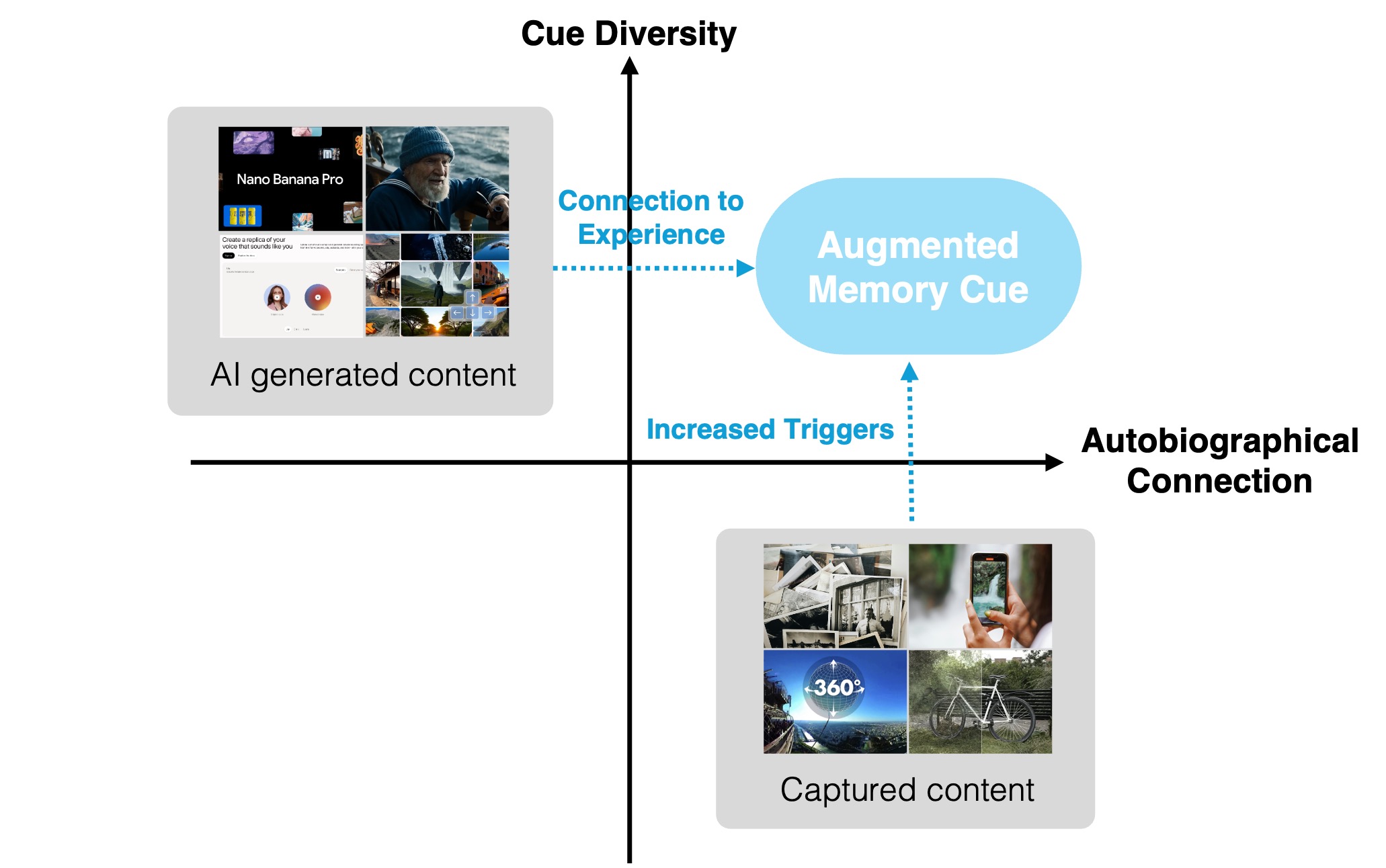}
\caption{Conceptual placement of \concept{} across autobiographical connection and cue diversity.}
\label{fig:scope}
\end{figure}

We introduce a novel category of memory retrieval cues, termed \concept{}. We define \concept{} as cues for memory recall that are grounded in captured media and extended through the generation of additional contextual information using generative AI. We identify two core dimensions of \concept{}: 1) autobiographical connection and 2) cue diversity (Figure~\ref{fig:scope}).

\begin{figure*}[t]
\centering
\includegraphics[width=\linewidth]{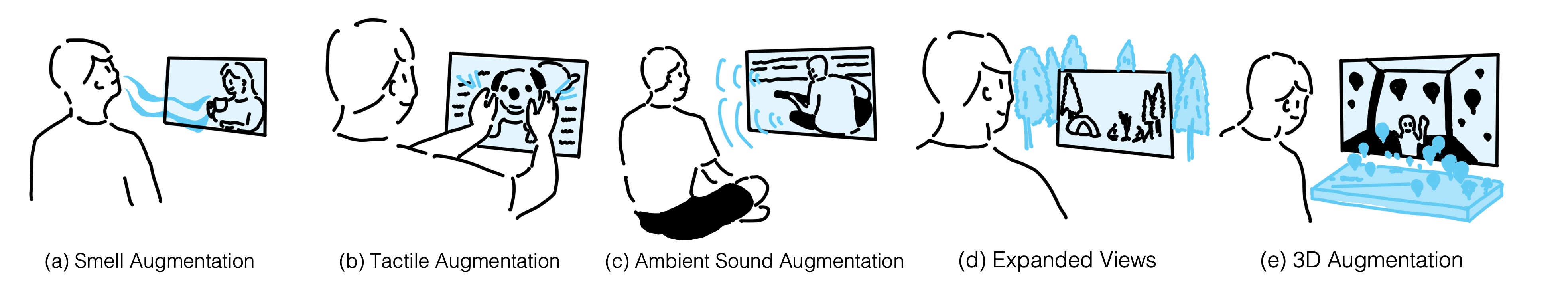}
\caption{Representative examples of \concept{}, illustrating sensory augmentations (smell (a), tactile (b), and ambient sound (c)) and visual-spatial extensions (expanded views (d) and 3D representations (e)).}
\label{fig:AugmentedMemoryCues}
\end{figure*}

\subsubsection*{\textbf{Autobiographical Connection}}

Autobiographical memory is shaped by what individuals perceived and experienced at the time of an event, including visual, emotional, and contextual information~\cite{conway2000construction,holland2010emotion}. Retrieval is typically facilitated when cues are congruent with the original encoding context, according to the encoding specificity principle~\cite{tulving1973encoding}. Prior work has shown such context effects across both physical environments~\cite{godden1975context} and linguistic contexts~\cite{marian2006language}. 
Therefore, \concept{} use captured personal media, such as photos and videos, as a scaffold for cue construction. Rather than generating entirely synthetic content, these cues are anchored in media originally captured by the user, which can help maintain a stronger link to the original lived experience. For example, generative AI can produce a scene based on a prompt such as “generate people in Paris.” In contrast, using a user’s own photograph containing people in Paris provides an experience-specific foundation. This grounding can strengthen autobiographical connection and may increase the likelihood of triggering memory recall.

\subsubsection*{\textbf{Cue Diversity}}
Different types of cues provide access to different aspects of the encoded experience. For example, olfactory cues can evoke autobiographical memories with different levels of emotional intensity~\cite{chu2002proust}, music can evoke vivid recollections~\cite{belfi2016music}, and bodily states can facilitate retrieval when congruent with the original experience~\cite{dijkstra2007body}. Therefore, increasing the diversity of cues can provide multiple, complementary pathways to access different aspects of the original experience.
Captured personal media such as photos and videos maintain a strong connection to lived experience, but they are inherently constrained by recording capabilities. A photograph provides only a fixed field of view and may exclude people or activities outside the frame. Similarly, discrete images do not represent what occurred between captured moments. \Concept{} address these limitations by leveraging generative AI to infer and extend contextual elements grounded in the captured content. For example, the system may generate contextually consistent movements of people entering or exiting the frame, thereby reconstructing dynamic aspects of the scene. Such extensions increase cue diversity and help convey additional contextual elements of the event, thereby providing additional cues that can support access to different aspects of the original experience.
\newline

Given these two core dimensions, we categorize existing media types according to their degree of autobiographical connection and cue diversity, as illustrated in Figure~\ref{fig:scope}. Captured media typically provide strong autobiographical grounding but limited cue diversity, while purely AI-generated content may offer diverse cues but lack direct experiential grounding. \Concept{} occupy the upper-right region of this conceptual space, combining strong autobiographical connection with expanded cue diversity through generative augmentation.

\subsection{Representative Forms}
\Concept{} are not limited to a single form. Instead, they can augment captured media in multiple ways while remaining grounded in the original experience. Figure~\ref{fig:AugmentedMemoryCues} illustrates five representative examples of \concept{}, spanning different ways in which captured media can be augmented.

These examples include sensory augmentations, such as smell, tactile, and ambient sound, as well as visual and spatial extensions, such as expanded views and 3D representations. For example, smell augmentation may evoke olfactory impressions associated with a photographed scene, such as the aroma of coffee in a cafe. Tactile augmentation may support recall by conveying implied physical properties, such as the texture of objects depicted in the scene. Ambient sound augmentation can further enrich recall by introducing contextually consistent environmental audio, such as waves, wind, music, or surrounding conversations. Expanded views present contextual elements beyond the original frame, such as surrounding trees in a campsite scene. 3D augmentation transforms captured media into spatial representations that reveal depth, layout, and relationships among elements.

Taken together, these examples illustrate the breadth of augmentations that \concept{} could support while remaining grounded in captured media. In this paper, we focus on one such form: dynamically augmented 3D memory dioramas. In Section~4.1, we motivate the diorama representation as a suitable medium for organizing and inspecting memory-relevant context, and in Section~4.2, we identify the recurring cue patterns that informed the design of our augmentation layers.

\begin{figure*}[t]
\centering
\includegraphics[width=\linewidth]{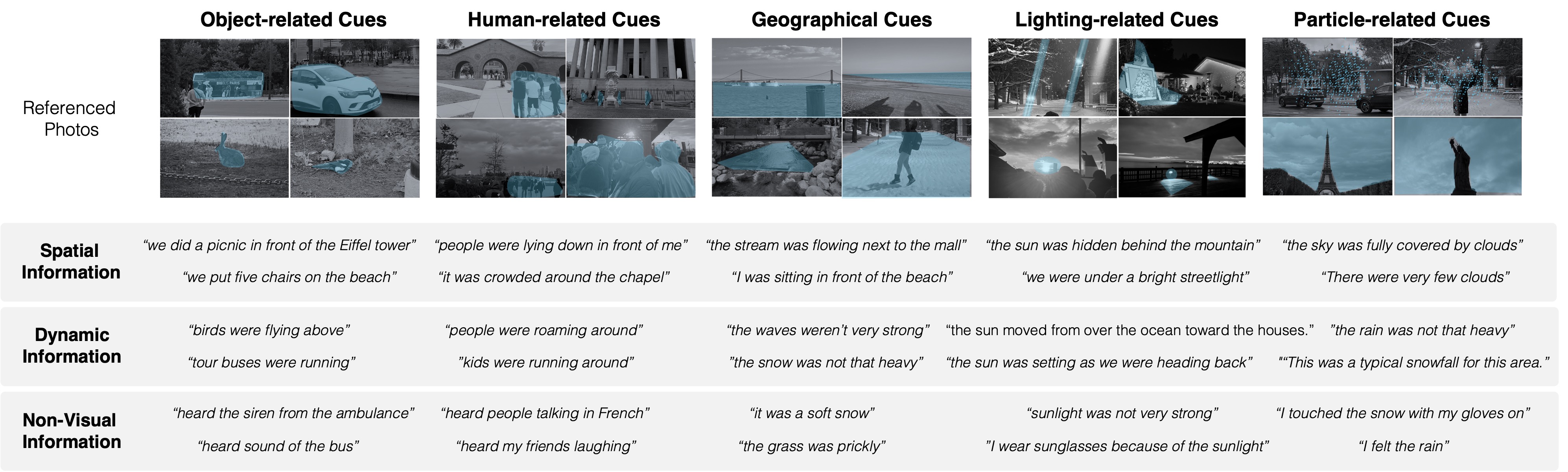}
\caption{Recurring cue patterns elicited from the formative study.}
\label{fig:formativestudy}
\end{figure*}
\section{\system{}: Design}

\system{} is a system that transforms personal photos into dynamic 3D dioramas, demonstrating the concept of \concept{}. In this section, we first describe the motivation for using a diorama representation, and then present a design elicitation study conducted to identify the essential components to support memory recall.

\subsection{Diorama Representation}
We adopt a 3D diorama representation for \system{}, motivated by prior work on the Worlds-in-Miniature (WIM) metaphor \cite{stoakley1995virtual}. Specifically, we leverage the following two features:

\subsubsection*{\textbf{Glanceable Overview:}}
A diorama can present multiple elements of a scene within a compact representation, allowing users to perceive the overall context at a glance. Prior work suggests that such representations support understanding of indoor floor layouts by showing the relative positions of multiple objects at once \cite{ibayashi2015dollhouse, stoakley1995virtual}, and can also help users understand larger-scale spatial layouts through the arrangement of buildings \cite{bluff2019don}. In the context of memory recall, this is important because memory cues are often distributed across various environmental elements. Presenting these diverse cues together within a unified space helps communicate not only individual details but also the overall atmosphere of the event.

\subsubsection*{\textbf{Multi-Angle Viewing:}}
Dioramas also support viewpoint-based inspection through small bodily movements around the diorama scene. Prior work suggests that complementary viewpoints can improve spatial awareness \cite{veas2012extended} and help users understand spatial relationships within an environment \cite{chittaro2005interactive}. In our context, this means that users can inspect the remembered scene from different angles, potentially revealing details and relationships that may be less apparent from a single fixed viewpoint.
\newline

Taken together, these properties make diorama representations well suited for \system{}, as they support the at-a-glance presentation of diverse cues and enable users to inspect remembered events from multiple perspectives.






\subsection{Formative Study}
To inform the design of the cue components of our system, we conducted a design elicitation study with eight participants. The study aimed to investigate how people recall past events and to identify the key components involved in the memory recall process.

\subsubsection{Participants}
We recruited eight participants from our local university community (4 male, 4 female). Each session lasted approximately 60 minutes. Participation was voluntary, and participants received no compensation.

\subsubsection{Method}
After obtaining informed consent, we explained the goals of the study and the overall procedure. Participants were first asked to select 5--10 photos taken at a single location and corresponding to one specific event. While viewing the selected photos, participants were instructed to recall the event in detail and describe the surrounding context (e.g., \textit{``people were walking in front of me,'' ``there was a large building ahead''}). They were encouraged to provide as much detail as possible, even if certain elements were not clearly visible in the photos. All sessions were audio-recorded for subsequent analysis. Two authors (A1 and A2) transcribed the audio recordings and conducted thematic coding to identify recurring types of recalled elements across participants.



\subsubsection{Elicited Cue Patterns}\label{cue-patterns}
From the study, we identified five recurring cue patterns that participants used when recalling past events from photos: 1) Object-related Cues, 2) Human-related Cues, 3) Geographical Cues, 4) Lighting-related Cues, and 5) Particle-related Cues (Figure~\ref{fig:formativestudy}). Importantly, these cue patterns were not recalled solely as static visual elements. Across categories, participants frequently referred to 1) spatial information, such as the locations and arrangements of scene elements, 2) dynamic information, such as movement, behavior, or temporal change, and 3) non-visual information, such as sound, temperature, and other bodily sensations.

\subsubsection*{\textbf{Object-related Cues}}
Participants frequently grounded their recollections in salient objects present in the scene, including vehicles (e.g., cars and buses) and animals (e.g., birds and rabbits) (P1, P2, P3, P4, P5). We categorized these object-centered cues as the \textit{Object-related Cues}. For instance, several participants (P1, P5) referred to birds visible in the photos and described remembering them flying overhead. In some cases, objects also triggered associated auditory memories. P1 recalled a tour bus and subsequently described the sound of its engine, while another remembered ambulances repeatedly passing by, accompanied by loud sirens. 

\subsubsection*{\textbf{Human-related Cues}}
Participants frequently recalled events in relation to the presence and activities of people, such as crowds at a location or companions during the event (P1, P2, P3, P6). We categorized these person-centered cues as the \textit{Human-related Cues}. Several participants described the spatial distribution and behaviors of people in the scene. For example, P1 recalled that many people were gathered throughout the park, with some having picnics and others lying on the grass. In addition, human-related cues often elicited sensory memories. P6 described feeling hot because the road was densely packed with people, recalling the body heat emanating from the crowd. Others linked people’s behavior to environmental conditions. For instance, P2 remembered people huddling together due to strong winds and cold weather. 

\subsubsection*{\textbf{Lighting-related Cues}}
Participants recalled events related to lighting conditions, including streetlights, decorative illumination, and sunlight (P5, P6). We categorized these illumination-based cues as the \textit{Lighting-related Cues}. When describing lighting, participants often elaborated on the spatial layout and atmosphere of the scene. For example, P6 recalled that the houses were aligned and brightly decorated, while the roadside area remained relatively dark. Such contrasts in illumination structured their memory of the environment.
Lighting cues were also linked to behavioral and experiential details. For instance, P5 remembered wearing sunglasses because the sunlight was too intense to comfortably read a book. 

\subsubsection*{\textbf{Geographical Cues}}
Participants recalled environmental features related to the landscape, such as sandy beaches, snow-covered ground, oceans, and rivers (P1, P4, P5, P7). We categorized these cues as the \textit{Geographical Cues}. In some cases, geographical elements were described in connection with weather conditions. For example, P5 noted that the ocean waves were relatively calm, recalling that it was not a windy day. Geographical cues also evoked haptic memories. P7 described the softness of the snow covering the ground, while P1 recalled the prickly sensation of touching grass. 

\subsubsection*{\textbf{Particle-related Cues}}
Participants recalled particle elements such as rain, snow, and clouds (P2, P7, P8). We categorized these cues as the \textit{Particle-related Cues}. 
In several cases, particle-related information was closely tied to affective experiences. For example, P2 described feeling cold and less motivated to walk due to cloudy weather. Participants also linked particle-related cues to observed human behavior. P8 recalled that people did not use umbrellas despite the rain, recalling the social practices in that location.

\begin{figure*}[t]
\centering
\includegraphics[width=\linewidth]{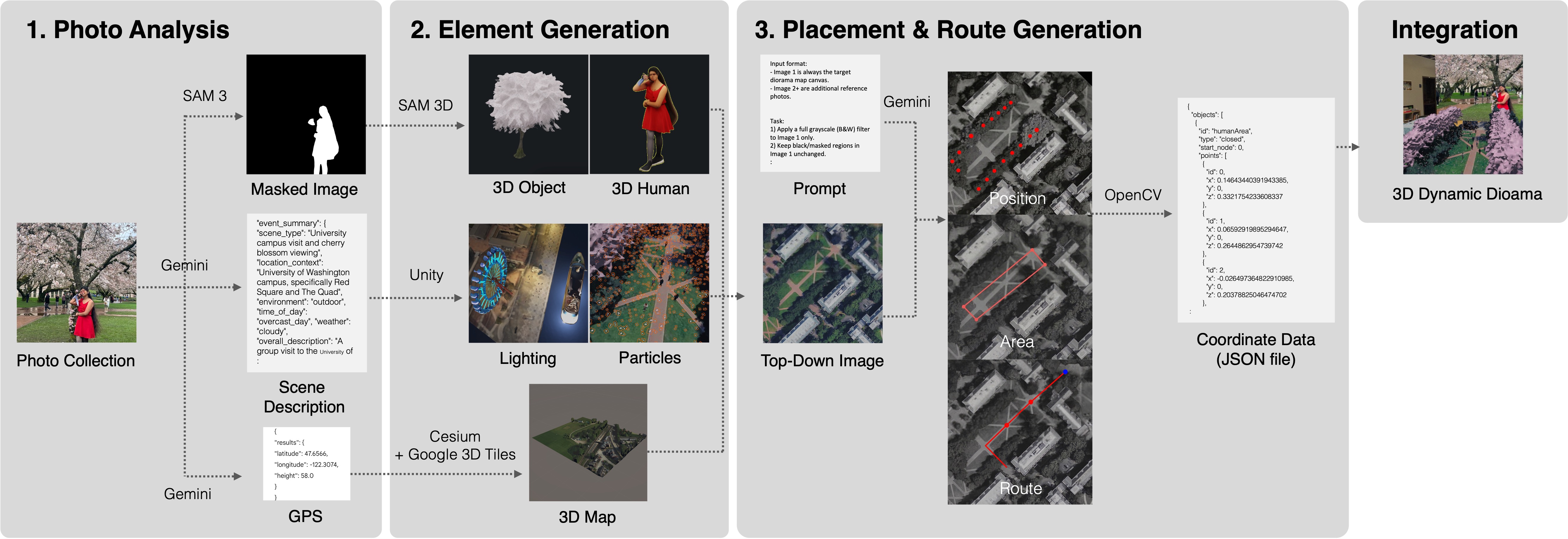}
\caption{\system{} system pipeline. The system transforms a photo collection from a single event into a dynamic 3D diorama through three phases: 1) photo analysis, 2) element generation, and 3) placement and route generation.}
\label{fig:systempipeline}
\end{figure*}
\section{\system{}: System}

\subsection{Overview}
In this section, we present \system{}, which employs a three-phase pipeline to transform input photos into dynamic 3D dioramas. Our formative study identified five recurring cue patterns in participants’ recollections: object-related, human-related, geographical, lighting-related, and particle-related cues. We translated these cue patterns into five corresponding augmentation layers in the system. Each layer is designed to reconstruct one type of contextual information that participants frequently drew on when recalling past events. While our formative study highlighted the potential of non-visual memory triggers (such as auditory and tactile elements), our current prototype focuses exclusively on visual augmentations. This scoping decision was made due to the inherent characteristics of the 3D diorama representation and current technical constraints in generating coherent multi-sensory stimuli.

\subsection{System Pipeline}
The pipeline consists of three phases: 1) photo analysis, 2) element generation, and 3) placement and route generation (Figure ~\ref{fig:systempipeline}). 
\subsubsection*{\textbf{Photo Analysis}}
In the photo analysis phase, the system extracts location metadata from the Exif data of the captured images, including latitude, longitude, and altitude. As a fallback mechanism, if Exif data is missing, the system can estimate the location using a vision-language model (e.g., Gemini 3~\cite{Gemini3}), although with reduced accuracy. Second, the system generates masked images from the input image using Segment Anything Model 3 (SAM 3) with a text prompt such as “person.” Third, the system extracts semantic scene descriptions. The system queries an LLM (Gemini 3) to generate detailed contextual descriptions and animation states for each masked element. To ensure accurate recognition, both the isolated masked image and the original unmasked image are provided to the LLM as prompts. 

\subsubsection*{\textbf{Element Generation}}
In the element generation phase, the system constructs the diorama scene through three primary steps: geographical grounding, 3D asset generation, and texture synthesis. First, the system generates the base 3D environment. We integrate Cesium, a geospatial plugin for Unity, with Google Photorealistic 3D Tiles. Using the aggregated location metadata obtained in the previous phase, the system automatically configures Cesium's georeferencing parameters to render a geographical foundation for the event.
Second, the system generates static 3D models for both objects and humans. We employ SAM 3D~\cite{chen2025sam} for image-to-3D conversion. Based on the event-level scene description JSON, the system retrieves the corresponding masked and unmasked image pairs and processes them through SAM 3D to construct individual 3D assets for each extracted element. Finally, the system synthesizes material textures. We leverage generative image models (Nano Banana 2~\cite{NanoBanana}) to create seamless textures for geographical surfaces and particle-based visual effects, enriching the visual appearance of the diorama.

\subsubsection*{\textbf{Placement and Route Generation}}
To determine the spatial placement and movement routes of the extracted elements, the system captures an orthographic top-down image of the generated 3D environment. This map image, along with the isolated image of each element, is provided to a multimodal LLM (Gemini 3 + Nano Banana 2). The LLM is tasked with outputting an annotated image that visually indicates the appropriate placements and routes. We adopted this approach because directly prompting the LLM to output numerical coordinates yielded spatially inconsistent results in our preliminary tests. To handle different types of elements, the system dynamically selects from three prompt templates: \textit{position} (e.g., for static objects), \textit{area} (e.g., for scattered environmental effects), and \textit{line} (e.g., for moving humans). 
The resulting annotated image is then processed using OpenCV. By applying strict color thresholding in the RGB color space, the system isolates the generated red pixels. 
Finally, the system projects these 2D pixel coordinates into the 3D scene using the orthographic top-down camera parameters and computes their intersections with the terrain mesh. These intersection points are then used as the 3D placement coordinates and route waypoints for each element.


\subsubsection*{\textbf{Integration and MR Deployment}}
In the final integration phase, the system aggregates all generated assets, including 3D meshes, synthesized textures, and the calculated spatial coordinates, into a unified scene within Unity. In our current implementation, these generated assets are imported into the Unity Editor, where the final scene is compiled. Within the Unity environment, the geographical foundation and the static 3D objects are composited based on the calculated coordinates. Dynamic environmental elements, such as lighting and particles, are overlaid based on the initial scene descriptions. Finally, the compiled scene is deployed to a standalone mixed reality headset (Meta Quest 3).

\subsection{Cue Component Layers}
\subsubsection{Object Layer}
\begin{figure}[h]
\centering
\includegraphics[width=\linewidth]{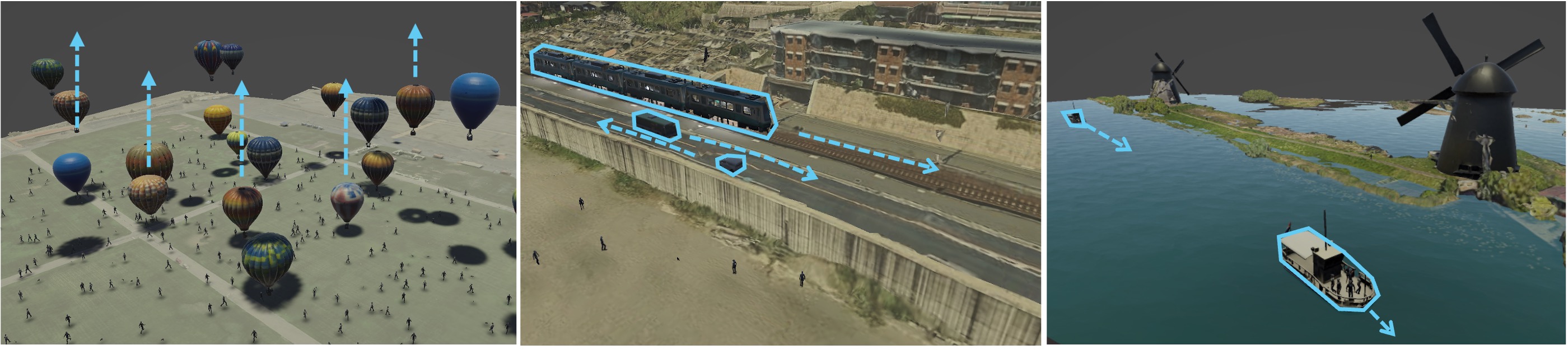}
\caption{Object layer. (a) Air balloon. (b) Train/car. (c) Boat.}
\label{fig:object-layer}
\end{figure}

In our formative study, participants frequently grounded their recollections in salient objects such as vehicles and animals, and often recalled their movement (Section~\ref{cue-patterns}). To support these cues, the object layer reconstructs salient scene objects and animates them (Figure~\ref{fig:object-layer}). We use 3D objects generated by SAM 3D. First, the system adjusts the scale of each object based on both the generated mesh size and the scene description of the object’s real-world size. Second, for static objects such as buildings and statues, the system places each generated 3D mesh according to the position description file. For moving objects, we consider two types of motion: vertical movement and horizontal movement. For vertical movement, objects move vertically according to the scene description, as in the case of air balloons and helicopters. For horizontal movement, including planes, cars, and trains, the system animates the objects along a spline defined by multiple control points derived from the generated route.

\subsubsection{Human Layer}
\begin{figure}[h]
\centering
\includegraphics[width=\linewidth]{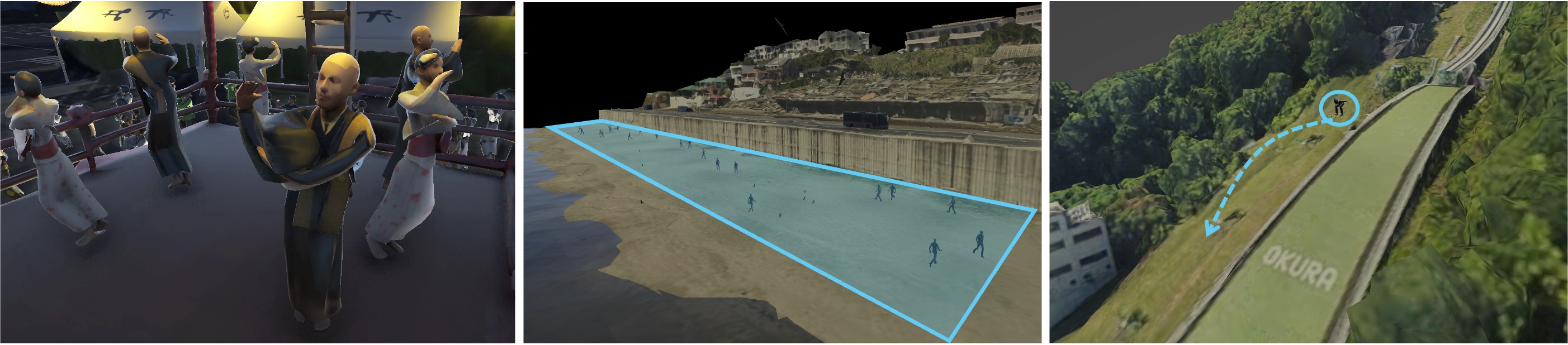}
\caption{Human layer. (a) Dancing. (b) Pedestrians walking. (c) Ski jumping.}
\label{fig:human-layer}
\end{figure}

Participants frequently recalled events in relation to the presence and behavior of people, such as crowds and companions (Section~\ref{cue-patterns}). To reconstruct these cues, the human layer populates the diorama with animated human figures (Figure~\ref{fig:human-layer}) and supports two types of placement. The first is placing pedestrians within a specified area. Using the spawning area from the area description file and the density from the scene description file, the system spawns pedestrians in that area. For this placement, we used pre-rigged human assets, as the generated human models from SAM 3D have no rig. The system assigns these pre-rigged models one of several predefined animations, including walking, running, and dancing, based on the scene description file. The second type is assigning specific movement to the human models generated from SAM 3D. Using the route description file, the system moves each human object along the specified route for motions such as ski-jumping and surfing. These models move along the route as static meshes.

\subsubsection{Particle Layer}
\begin{figure}[h]
\centering
\includegraphics[width=\linewidth]{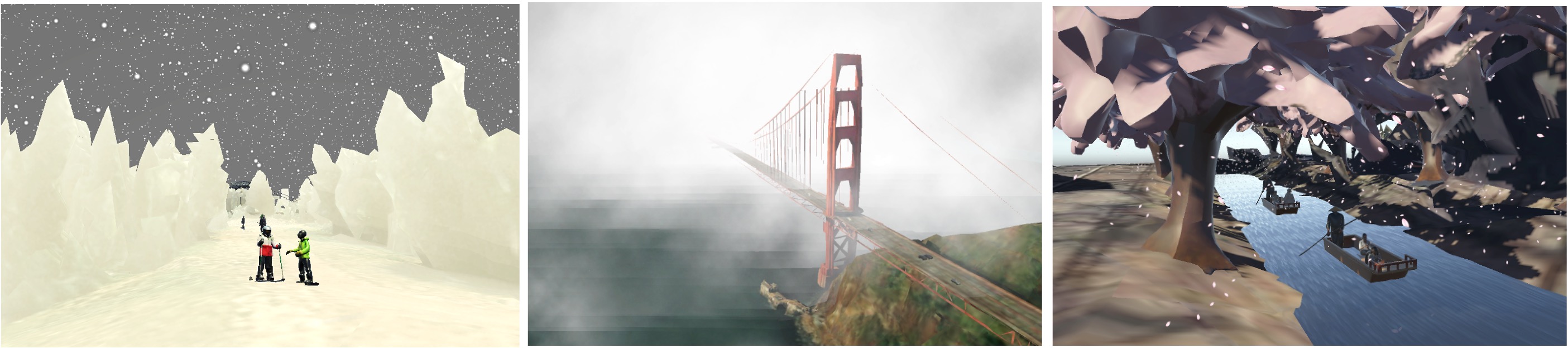}
\caption{Particle layer. (a) Snow. (b) Fog. (c) Cherry blossom.}
\label{fig:particle-layer}
\end{figure}

Participants recalled atmospheric elements such as rain, snow, and clouds, often linking them to their emotions such as feeling unmotivated (Section~\ref{cue-patterns}). To convey these ambient cues, the particle layer overlays environmental particle effects onto the diorama scene (Figure~\ref{fig:particle-layer}). The system includes several predefined particle types, including snow, rain, and fog. These particle effects are implemented in Unity using the particle system. For each of these effects, the system provides three intensity levels: low, medium, and high. Based on the scene description file, the system activates the corresponding effects. In addition, the system supports other particle effects using generated textures, such as falling leaves and cherry blossoms. The system uses the generated texture image and applies it to the material of the particle system.

\subsubsection{Lighting Layer}
\begin{figure}[h]
\centering
\includegraphics[width=\linewidth]{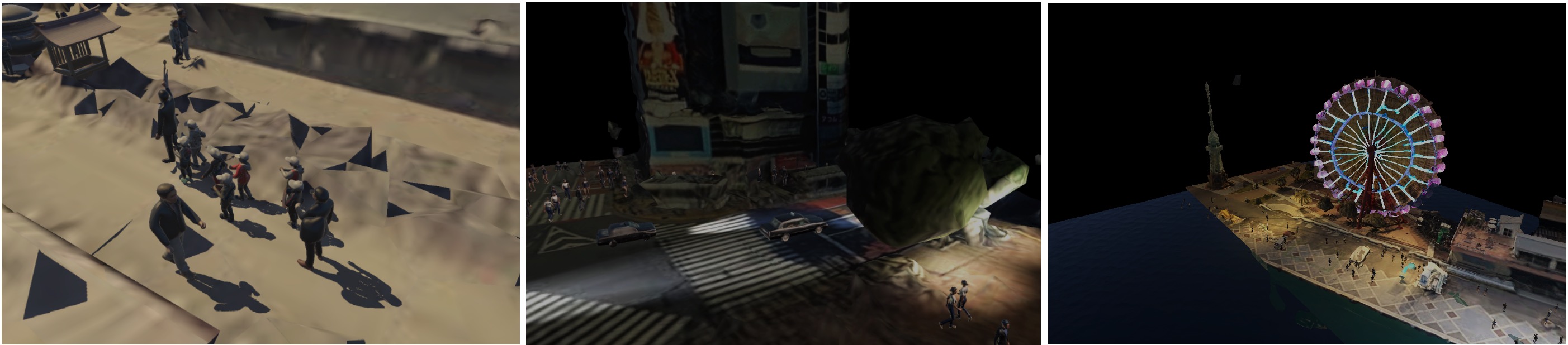}
\caption{Lighting layer. (a) Sunset lighting. (b) Streetlight. (c) Illuminated objects.}
\label{fig:lighting-layer}
\end{figure}

Participants recalled lighting conditions as structuring their memory of the environment, such as brightly decorated houses against dark roadsides (Section~\ref{cue-patterns}). To recreate these illumination cues, the lighting layer supports three types of lighting effects: sunlight, streetlights, and illuminated objects (Figure~\ref{fig:lighting-layer}). Sunlight changes the overall lighting of the scene and creates shadows, thereby recreating the mood of the event. Based on the time specified in the scene description file, the system adjusts the rotation of the directional light in Unity, changing the overall brightness and lighting direction of the diorama. Streetlights are implemented using spot lights in Unity, and their positions are determined based on the position description file. Illuminated objects support lighting effects for objects such as Ferris wheels and projected decorations on buildings in amusement parks. These effects are implemented by applying custom emissive textures to the corresponding objects in Unity.

\subsubsection{Geographical Layer}
\begin{figure}[h]
\centering
\includegraphics[width=\linewidth]{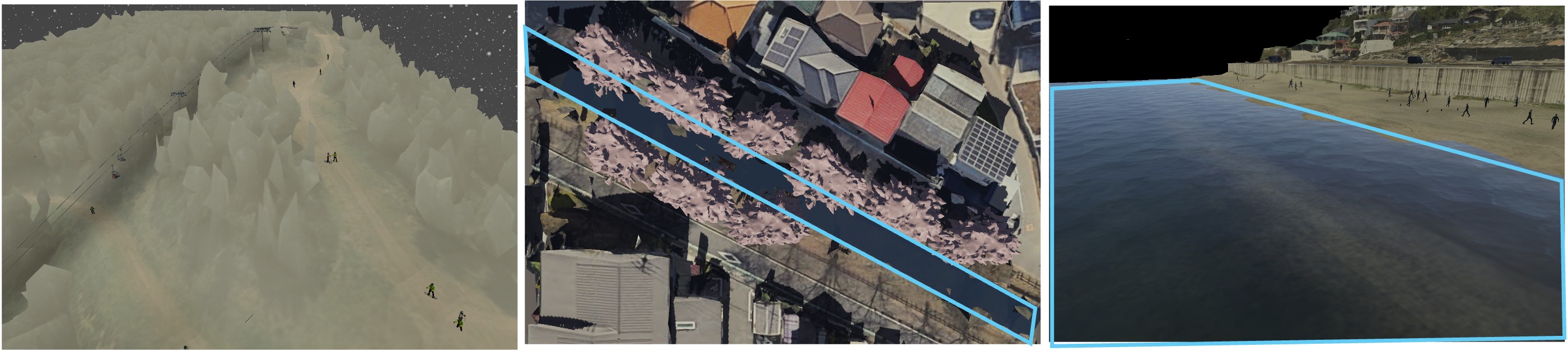}
\caption{Geographical layer. (a) Snow-covered terrain. (b) Animated river. (c) Animated ocean.}
\label{fig:geographical-layer}
\end{figure}

Participants recalled environmental features such as beaches, snow-covered ground, and water bodies, connecting them to weather conditions or tactile sensations (Section~\ref{cue-patterns}). To reconstruct these location-specific elements, the geographical layer augments the base map with two types of effects: terrain texture effects and water effects (Figure~\ref{fig:geographical-layer}). For terrain texture effects, the system uses generated texture images and applies them to the 3D map texture, changing the appearance of the terrain according to the scene description, such as to a snowy landscape or a grass-covered surface. For water effects, the system uses the area description file to identify water regions, such as oceans and rivers. It then overlays these regions with animated water textures with wave effects to represent water movement.

\subsection{Implementation Details}
The processing pipeline was executed on an NVIDIA DGX Spark (Cortex-X925 / Cortex-A725 CPU, 128 GB RAM, NVIDIA GB10 GPU) running Ubuntu 24.04.3 LTS. The data processing scripts and color thresholding pipelines were implemented in Python (3.12.3) using OpenCV (4.13.0). The MR application was developed using a MacBook Pro (M4 Max with 16-Core CPU, 128GB RAM) with Unity 6000.1.15f1. To render the geographical foundation, we integrated the Cesium for Unity (v1.19.0) plugin with Google Photorealistic 3D Tiles. The final diorama scene was built and deployed to the Meta Quest 3 using the Meta XR All-in-One SDK (v81.0.0).

\subsection{Technical Feasibility Test}
\subsubsection{Method}
We conducted a technical feasibility test to assess the computational performance and end-to-end feasibility of the pipeline. We used 25 photo sets from three authors (A1, A2, and A3), each corresponding to a single event at a single location and consisting of five photos. Because autobiographical consistency and scene context can only be judged by the original photographer, each author assessed the dioramas generated. The formal evaluation of \system{}'s effect on human memory recall is presented in our user study (Section~\ref{section:user-study}). We evaluated the three phases separately. 

In photo analysis, a failure was defined as a scene description that was clearly inconsistent with the photographed event. In element generation, a failure was defined as an implausible generated element. In placement and route generation, a failure was defined as a missing or semantically incorrect placement or motion path. We also measured the runtime of each stage in the pipeline.

\subsubsection{Results}
The average end-to-end processing time per diorama was 15.70 minutes. The success rates were: 1) photo analysis: 100\%, 2) element generation: 92.58\%, and 3) placement and route generation: 75.25\%. Typical failures in element generation involved implausible outputs in which multiple objects were merged into a single object, such as an animal with two heads. Typical failures in placement/route generation mainly stemmed from errors in generating the annotated top-down image. These failures were primarily of two types: producing incorrectly oriented images, and placing annotations on irrelevant generated images.


\begin{table}[h]
\centering
\caption{Technical feasibility results for each pipeline stage.}
\label{tab:technical_evaluation}
\footnotesize
\begin{tabular}{lccc}
\hline
\textbf{Stage} & \textbf{Time (min)} & \textbf{Count} & \textbf{Success Rate (\%)} \\
\hline
Photo Analysis & 0.29 & 1.00 & 100.00 \\
Element Generation & 10.96 & 10.24 & 92.58 \\
Placement/Route Generation & 4.45 & 11.80 & 75.25 \\
\hline
End-to-End & 15.70 & -- & -- \\
\hline
\end{tabular}
\end{table}


\begin{figure*}[t]
\centering
\includegraphics[width=\linewidth]{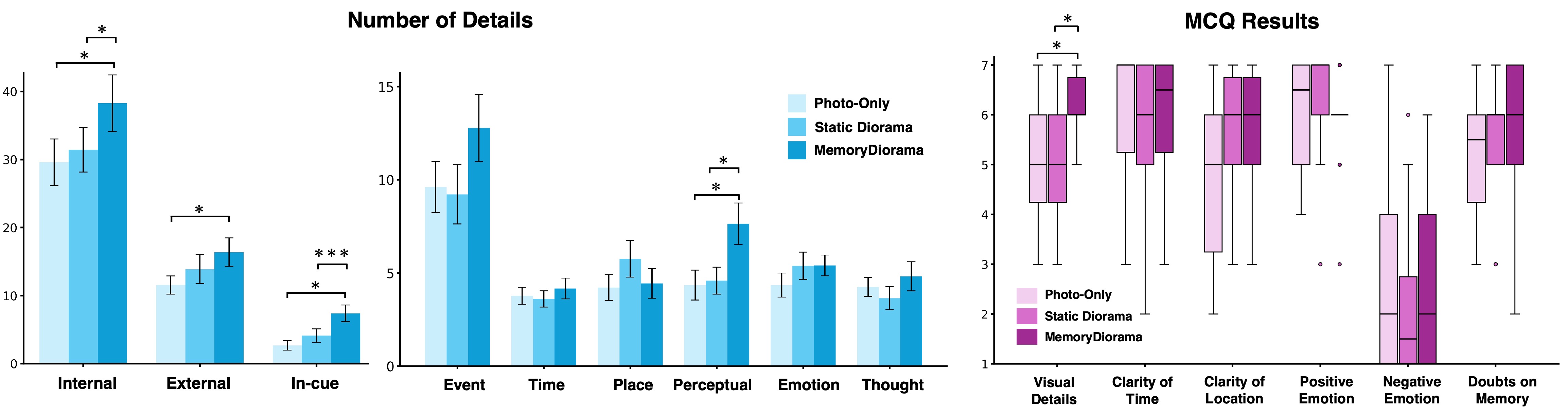}
\caption{
User study results comparing Photo-Only, Static Diorama, and \system{}. Left and middle panels show mean detail counts with standard error bars; the right panel shows box plots of questionnaire responses (*: p<.05, **: p<.01, ***: p<.001).
}
\label{fig:userstudy-results}
\end{figure*}
\section{User Study}\label{section:user-study}
We conducted a within-subject user study with 18 participants to evaluate the effects of \system{} on autobiographical memory recall and user experience. 
\subsection{Method}
\subsubsection{Participants}
We recruited 18 participants (9 male, 9 female; age = 21--36 years, M = 26.17, SD = 3.37) from a local university community. Each session lasted approximately 40 minutes, and participants received \$15 as compensation.

\subsubsection{Conditions}
We employed a within-subject design with three conditions to evaluate the effectiveness of our system. To control for the novelty effect of MR, all conditions were conducted in an MR environment using a Quest 3 headset. The conditions were:

\begin{itemize}
    \item \textbf{Photo-Only:} 2D photos were displayed on a virtual screen shown in front of the participant. 
    \item \textbf{Static Diorama:} In addition to the 2D photos on the front screen, a static 3D diorama generated from Google Photorealistic 3D Tiles was displayed below it.
    \item \textbf{\system{}:} Participants viewed the 2D photos along with a dynamic 3D diorama.
\end{itemize}

Each participant provided a pool of photo collections before the study, where each collection corresponded to a distinct autobiographical event and consisted of five photos. For each candidate event, participants reported its significance and memory age. Following prior work~\cite{martin2022smartphone}, we first excluded candidate events with low significance ratings and low ratings for how well the events were remembered (ratings of 1 or 2 on a 5-point scale). We then selected the final three events by minimizing differences in how long ago they had occurred. Collections referring to the same trip/day/occasion were not assigned across different conditions. In all conditions, the five photos were displayed in front of the participant. 

\subsubsection{Procedure}
After obtaining informed consent, participants received a brief walkthrough of the three conditions and were given time to familiarize themselves with the MR environment.
To assess autobiographical memory recall, we adopted an Autobiographical Interview–based procedure~\cite{levine2002aging}. In each trial, participants were asked to verbally recall memories associated with one assigned photo collection. They were instructed to recall all memories related to the event and to provide as much detail as possible. No additional prompts were provided during the initial narration. After the initial recall, a general probe was administered to allow further elaboration. Initial and probed responses were combined for analysis. All narrations were audio-recorded.

Each participant completed three recall trials (one trial per condition). A five-minute break was provided between conditions. We fully counterbalanced condition order across participants. After each trial, participants completed post-condition questionnaires. 

\subsubsection{Measurements}
Audio recordings were transcribed and segmented into discrete informational details. Following established autobiographical interview procedures~\cite{levine2002aging, renoult2020classification, martin2022smartphone}, each detail was classified as internal, external, or in-cue. Internal details reflect event-specific episodic re-experiencing. External details include semantic information, repetitions, or off-target events. In-cue details describe information about the cues themselves. The transcripts were initially divided between two raters (A1 and A2) for coding. To assess inter-rater reliability, 9 events were randomly selected for double coding by the other rater. Intraclass correlation coefficients for internal, external, and in-cue detail counts all exceeded 0.90, indicating high inter-rater agreement.

After each recall task, participants rated the recalled memories on a 7-point Likert scale using selected items from the Memory Characteristics Questionnaire (MCQ)
~\cite{johnson1988phenomenal, d2003phenomenal}. The items assessed: 1) visual details (1 = none, 7 = a lot), 2) clarity of location (1 = not at all clear, 7 = very clear), 3) clarity of time (1 = not at all clear, 7 = very clear), 4) positive emotion (1 = none, 7 = very intense), 5) negative emotion (1 = none, 7 = very intense), 6) doubts about memory accuracy (1 = a great deal of doubt, 7 = no doubt whatsoever).

Workload was assessed using the NASA Task Load Index (NASA-TLX)~\cite{hart1988development}, and enjoyment was measured with a 7-point Likert item (“I enjoyed using this system”).





\subsection{Results}
For all outcome measures, the absolute skewness and kurtosis values were below 3.0 and 10.0, respectively, indicating that normality was considered acceptable~\cite{kline2023principles, hwang2025telepulse}. 
We therefore conducted one-way repeated-measures ANOVAs across all outcome measures, followed by Holm-corrected paired-samples t-tests for significant main effects. The significance threshold was set at $\alpha = .05$.

\subsubsection{Number of Recalled Details}
Figure~\ref{fig:userstudy-results} (left) shows the number of the recalled memory details. For \textbf{internal details}, ANOVA showed a significant effect of condition, $F(2, 34) = 5.50$, $p = .0085$, $\eta_p^2 = .24$. Post hoc tests showed that the \textit{\system{}} ($M = 38.28$, $SD = 17.66$) elicited more internal details than \textit{photo-only} ($M = 29.61$, $SD = 14.52$; $p = .0449$, $d_z = 0.64$) and \textit{static diorama} ($M = 31.44$, $SD = 13.92$; $p = .0465$, $d_z = 0.59$). For \textbf{external details}, ANOVA also showed a significant effect of condition, $F(2, 34) = 3.84$, $p = .0313$, $\eta_p^2 = .18$. Post hoc tests showed that \textit{\system{}} ($M = 16.39$, $SD = 8.89$) elicited more external details than \textit{photo-only} ($M = 11.56$, $SD = 5.67$; $p = .0289$, $d_z = 0.69$), whereas \textit{static diorama} fell in between ($M = 13.89$, $SD = 9.00$) and did not differ significantly from either condition. For \textbf{in-cue details}, ANOVA showed a significant effect of condition, $F(2, 34) = 10.32$, $p = .0003$, $\eta_p^2 = .38$. Post hoc tests showed that \textit{\system{}} ($M = 7.39$, $SD = 5.17$) elicited more in-cue details than \textit{photo-only} ($M = 2.67$, $SD = 2.91$; $p = .0056$, $d_z = 0.82$) and \textit{static diorama} ($M = 4.11$, $SD = 4.21$; $p = .0007$, $d_z = 1.10$). 

Among the \textbf{internal detail subcategories}, only perceptual details showed a condition effect. ANOVA revealed a significant effect of condition, $F(2, 34) = 5.98$, $p = .0059$, $\eta_p^2 = .26$. Post hoc tests showed that the \system{} condition ($M = 7.22$, $SD = 4.80$) elicited more perceptual details than both \textit{photo-only} ($M = 4.11$, $SD = 3.36$; $p = .0381$, $d_z = 0.66$) and \textit{static diorama} ($M = 4.33$, $SD = 3.09$; $p = .0381$, $d_z = 0.65$). 

For the \textbf{ratio of internal details} to total recalled details, ANOVA showed no significant effect of condition, $F(2, 34) = 0.85$, $p = .4356$, $\eta_p^2 = .05$. Post hoc tests also showed no significant pairwise differences between \textit{photo-only} ($M = 0.6641$, $SD = 0.1338$), \textit{static diorama} ($M = 0.6566$, $SD = 0.1357$), and \textit{\system{}} ($M = 0.6168$, $SD = 0.1395$).

\subsubsection{Subjective Memory Characteristics Results}
Figure~\ref{fig:userstudy-results} (right) shows the results of the Memory Characteristics Questionnaire. Only \textbf{visual details} showed a significant effect of condition. ANOVA revealed a significant effect, $F(2, 34) = 4.59$, $p = .0172$, $\eta_p^2 = .21$. Post hoc tests showed that the \textit{\system{}} ($M = 6.06$, $SD = 0.73$) was rated higher than both \textit{photo-only} ($M = 5.17$, $SD = 1.25$; $p = .0403$, $d_z = 0.65$) and \textit{static diorama} ($M = 5.22$, $SD = 1.26$; $p = .0408$, $d_z = 0.60$). 

\subsubsection{User Experience}
For the \textbf{enjoyment} ratings, ANOVA showed a significant effect of condition, $F(2, 34) = 10.92$, $p = .0002$, $\eta_p^2 = .39$. Post hoc tests showed that the \system{} ($M = 6.33$, $SD = 0.77$) was rated higher than both \textit{photo-only} ($M = 4.89$, $SD = 1.37$; $p = .0006$, $d_z = 1.12$) and \textit{static diorama} condition ($M = 5.61$, $SD = 1.24$; $p = .0158$, $d_z = 0.71$). For the \textbf{NASA-TLX} average score (0--100 scale), ANOVA showed no significant effect of condition, $F(2, 34) = 0.46$, $p = .6368$, $\eta_p^2 = .03$. Mean NASA-TLX scores were 9.78 ($SD = 9.37$) in \textit{photo-only}, 12.39 ($SD = 13.16$) in \textit{static diorama}, and 10.06 ($SD = 8.41$) in \textit{\system{}}.

\section{Discussion and Future Work}
\subsubsection*{\textbf{\system{} Increased Internal Details}}
As expected, \system{} increased in-cue details, likely because the dynamic diorama presented more visual and contextual information than the photo-only and static diorama conditions. Notably, \system{} also increased the number of internal details. This pattern suggests that the richer presentation was associated with broader recollection of the remembered events. Further analyses provide a more nuanced view of this effect. In the subcategory analysis, \system{} increased perceptual details, indicating that participants verbalized more sensory aspects of their experiences. For example, dynamic cues such as ocean movement and atmospheric particles were accompanied by recollections of cold and wind, including \textit{``it was pretty cold by the oceanfront''} (P14) and \textit{``It got windy suddenly''} (P10). Cues such as sunlight were also associated with recollections of heat, as one participant described the place as \textit{``it was like a wooden sauna''} (P11). Questionnaire responses also showed higher visual-detail ratings under \system{}, suggesting that participants experienced the remembered episodes as more visually vivid. Together, these findings suggest that the expanded contextual cues in \system{} did not merely provide more content to describe, but may have scaffolded richer sensory and perceptual recollection.

At the same time, \system{} did not increase the proportion of internal details relative to the total number of recalled details. This suggests that the system did not selectively enhance memory specificity. Rather, it appeared to support a broader increase in recalled content, potentially driven by both richer recollective experience and the greater amount of available cue information.





\subsubsection*{\textbf{Potential for False Memory Generation}}
False memories have been widely studied in cognitive science, defined as memories of events that never occurred or that are remembered differently from how they actually occurred~\cite{roediger1995creating}. False memories can also be created for autobiographical events. For example, participants have been led to recall childhood events that never occurred through exposure to false narratives~\cite{loftus1995formation, garry2005actually}, narratives accompanied by authentic photographs~\cite{lindsay2004true}, and digitally manipulated photographs~\cite{wade2002picture}.

In \system{}, we generate \concept{} from captured content. Because these cues are generated through predictive processes, they may contain inaccuracies or misinformation. In our user study, we did not have access to participants’ full autobiographical experiences beyond the captured content. Therefore, we could not determine the extent to which misinformation was introduced or whether it contributed to the formation of false memories. Future work should systematically investigate the potential impact of our system on false memory formation.

\subsubsection*{\textbf{Limitations of the User Study.}} Our user study has several limitations. First, although we used a within-subject design and matched events by memory significance and age, each condition was evaluated with a different autobiographical event. As a result, condition effects cannot be fully separated from event-level differences in memorability or narratability. Second, our comparison evaluates the effect of the full \system{} system rather than isolating the contribution of individual cue layers. Third, although we presented all conditions in MR to control for headset novelty and display format, viewing personal photos in MR is not a common everyday reminiscence practice. Finally, our study did not assess memory accuracy or false-memory formation. Therefore, further research is needed to clarify the generalizability, individual cue effects, and risks of false-memory formation.



\subsubsection*{\textbf{Augmented Memory Cues Across Modalities}}
Research in cognitive psychology has shown that different sensory modalities (e.g., smell, sound, and vision) exhibit distinct characteristics in autobiographical memory retrieval. For example, while verbal and visual cues tend to evoke memories from early adulthood (11--20 years), odor cues more often trigger memories from the first decade of life (<10 years)~\cite{willander2006smell}.
In addition, odor cues have been found to differ qualitatively from visual cues and to yield a greater amount of autobiographical detail during recall~\cite{chu2002proust}.

In this study, our focus was primarily on visual cues. However, the concept of \concept{} could also be extended to other sensory modalities, including olfaction and haptics, based on previous work such as \textit{Smell \& Paste}~\cite{brooks2023smell} and \textit{Scene2Hap}~\cite{jingu2025scene2hap}. Such augmented cues may either amplify or attenuate the characteristic properties of each modality in memory retrieval. Future work should examine how \concept{} influence memory retrieval across different sensory modalities.

\section{Conclusion}
We introduced \concept{}, a concept that extends captured personal media with generative AI to support autobiographical memory recall. Based on this concept, we presented \system{}, a system that transforms everyday photos into dynamic 3D dioramas. Our user study showed that \system{} elicited more internal and in-cue details than both Photo-Only and Static Diorama. It also increased perceptual details, visual vividness, and enjoyment, suggesting a richer recollective experience, while maintaining comparable workload. At the same time, \system{} did not increase the proportion of internal details relative to total recalled details, indicating that it supported a broader increase in recollection rather than selectively enhancing episodic specificity. These findings suggest that \concept{} can support autobiographical memory recall and open new opportunities for designing immersive and reflective memory systems.

\ifdouble
  \balance
\fi
\bibliographystyle{ACM-Reference-Format}
\bibliography{references}
\section*{Appendix}
In this appendix, we present the prompts used in our system pipeline, including prompts for scene analysis, texture generation, and spatial layout generation.

\subsection{Scene Analysis Prompts}

\begin{lstlisting}[label={lst:particles_json}]
Your task is to analyze a small set of photos from a single real-world event and convert them into structured scene descriptions.

You will receive:
1. One or more original photos from the same event.
2. Optional masked crops for specific elements extracted from the photos.
3. Optional metadata such as timestamp, GPS, latitude, longitude, altitude, or location estimates.

Your goal is not to produce a generic caption.
Your goal is to infer scene elements that are useful for reconstructing a dynamic 3D diorama for memory recall.

The output must be grounded in visible evidence.
You may make conservative contextual inferences only when they are strongly supported by the images.
Do not invent salient elements that are not visible or strongly implied.

Analyze the scene using the following five cue layers:
1. object-related cues
2. human-related cues
3. geographical cues
4. lighting-related cues
5. particle-related cues

For each detected element, infer:
- semantic label
- concise physical description
- likely real-world role in the scene
- motion or animation state if applicable
- approximate size category
- confidence score
- which image(s) support the inference

Important rules:
- Use the masked image to identify the target element precisely.
- Use the original image to recover context, scale, and surroundings.
- Prefer concrete nouns and observable actions.
- Keep descriptions short and operational for downstream generation.
- Avoid subjective interpretation unless it directly affects reconstruction, such as "crowded", "sunset", "light rain".
- Do not mention camera properties, composition, or photographic style unless they indicate actual environmental conditions.
- If uncertain, use "unknown" or lower confidence rather than guessing.
- If an element is static, set animation to "static".
- If an element is dynamic, choose a simple verb phrase such as "walking", "running", "flying", "driving", "floating", "waving", "falling", "flowing".
- Infer weather or atmosphere only if visible evidence exists, such as clouds, snow, rain, fog, wet ground, haze, blowing trees, umbrellas.
- Infer lighting from visible environmental evidence, such as sun direction, shadows, sunset tones, streetlights, decorative illumination, reflections, dark roadsides.

Output format:
Return a single valid JSON object only.
No markdown.
No explanations outside JSON.

JSON schema:

{
  "event_summary": {
    "scene_type": "string",
    "location_context": "string",
    "environment": "indoor | outdoor | mixed | unknown",
    "time_of_day": "day | night | sunset | sunrise | overcast_day | unknown",
    "weather": "clear | cloudy | rainy | snowy | foggy | windy | mixed | unknown",
    "overall_description": "string"
  },
  "objects": {
    "object01": {
      "images": ["string"],
      "label": "string",
      "description": "string",
      "animation": "string",
      "size": "small | medium | large | unknown",
      "confidence": 0.0
    }
  },
  "humans": {
    "human01": {
      "images": ["string"],
      "count_type": "individual | group",
      "description": "string",
      "animation": "string",
      "pose_or_activity": "string",
      "confidence": 0.0
    }
  },
  "geography": {
    "geo01": {
      "images": ["string"],
      "type": "road | beach | river | ocean | mountain | snowfield | grass | urban_block | park | bridge | station | other",
      "description": "string",
      "dynamic_state": "static | flowing | waving | unknown",
      "confidence": 0.0
    }
  },
  "lighting": {
    "light01": {
      "images": ["string"],
      "type": "sunlight | sunset | streetlight | decorative_light | indoor_light | overcast_light | unknown",
      "description": "string",
      "intensity": "low | medium | high | unknown",
      "direction_or_area": "string",
      "confidence": 0.0
    }
  },
  "particles": {
    "particle01": {
      "images": ["string"],
      "type": "rain | snow | fog | cloud | mist | falling_leaves | blossoms | none | unknown",
      "description": "string",
      "intensity": "low | medium | high | unknown",
      "confidence": 0.0
    }
  }
}

Additional constraints:
- Include only elements relevant for diorama reconstruction.
- Merge duplicates across images when they refer to the same semantic element type and event context.
- If no element is found for a layer, return an empty object for that layer.
- Keep the JSON compact and deterministic.
- Use English for all labels and values.
\end{lstlisting}

\subsection{GPS Prediction Prompts}
\begin{lstlisting}[label={lst:particles_json}]
You are an assistant that estimates one single GPS location from multiple images.

Analyze all provided images together and infer one best shared location.

Use visual cues such as terrain, roads, buildings, vegetation, coastline, mountains, and overall landscape context.
Also estimate elevation in meters.

Input images:
{joined}

Return JSON only.
Do not return markdown.
Do not add explanations.
Do not add extra keys.
Use exactly this format:

{{
  "results": {{
    "latitude": 0.0,
    "longitude": 0.0,
    "height": 0.0
  }}
}}
\end{lstlisting}

\subsection{Geographical Texture Generation}
\begin{lstlisting}[label={lst:particles_json}]
Generate a stylized top-down surface-cover texture for the geographical layer of a miniature memory diorama.

Scene context:
{scene_summary}

Surface cover type:
{surface_cover_type}

Visual characteristics:
{surface_cover_description}

Style requirements:
- the texture should cover the overall terrain surface of the diorama
- top-down view
- stylized, soft, and visually cohesive with a handcrafted memory diorama aesthetic
- preserve the visual identity of the surface cover material
- suitable for overlaying across the terrain as a global geographical texture
- visually readable at small scale
- no text, labels, icons, figures, buildings, or perspective objects
- no strong directional shadows or dramatic lighting
- texture only

Output a single square texture image.
\end{lstlisting}

\subsection{Particle Texture Generation}
\begin{lstlisting}[label={lst:particles_json}]
Generate a stylized texture asset for animated environmental particles in a miniature memory diorama.

Scene context:
{scene_summary}

Particle type:
{particle_type}

Desired motion impression:
{motion_description}

Style requirements:
- soft, lightweight, semi-transparent appearance
- visually simple and clean
- suitable for repeated use as a particle texture in animation
- consistent with a handcrafted, memory-inspired diorama style
- no background scene
- isolated texture asset only
- no text, symbols, or decorative framing
- should remain readable when duplicated many times in motion

Output a single square texture image with a plain clean background or transparent-style appearance.
\end{lstlisting}

\subsection{Position Generation}
\begin{lstlisting}[label={lst:particles_json}]
Input format:
- Image 1 is always the target diorama map canvas.
- Image 2+ are additional reference photos.

Task:
1) Apply a grayscale (B&W) filter to Image 1 only.
2) Keep black/masked regions in Image 1 unchanged.
3) Draw exactly one solid red vertex (dot) at the center location of {object} on Image 1.

Hard constraints:
- Draw only on Image 1. Do not draw on any reference images.
- Draw exactly one red vertex (#FF0000). No lines, labels, or extra marks.
- Keep Image 1 framing and resolution unchanged.
- Return one edited image based on Image 1.
\end{lstlisting}

\subsection{Area Generation}
\begin{lstlisting}[label={lst:particles_json}]

Input format:
- Image 1 is always the target diorama map canvas.
- Image 2+ are additional reference photos.

Task:
1) Apply a full grayscale (B&W) filter to Image 1 only.
2) Keep black/masked regions in Image 1 unchanged.
3) On Image 1, draw one closed polygon for the {object} area:
   - Red vertices (#FF0000)
   - Red connecting lines (#FF0000)

Hard constraints:
- Draw only on Image 1. Do not draw on any reference images.
- Draw exactly one closed polygon; no open lines and no extra polygons.
- Use only red lines and red vertices. No labels or extra marks.
- Keep Image 1 framing and resolution unchanged.
- Return one edited image based on Image 1.
\end{lstlisting}
\subsection{Route Generation}
\begin{lstlisting}[label={lst:particles_json}]
Input format:
- Image 1 is always the target diorama map canvas.
- Image 2+ are additional reference photos.

Task:
1) Apply a grayscale (B&W) filter to Image 1 only.
2) Overlay one movement path for {object} on Image 1.

Strict drawing rules:
- Use ONLY Red #FF0000 for the path and red vertices.
- Use ONLY Blue #0000FF for the start marker.
- Path line width: 8 pixels.
- Place one blue dot at the start, overlapping the first red path segment.
- Add red vertices at key turning or altitude-change points.
- Draw exactly one simple connected non-loop path (no forks, no branches).

Hard constraints:
- Draw only on Image 1. Do not draw on any reference images.
- No gradients, shadows, transparency, labels, or extra marks.
- Keep Image 1 framing and resolution unchanged.
- Return one edited image based on Image 1.
\end{lstlisting}
\subsection{Particle System Generation}
\begin{lstlisting}[label={lst:particles_json}]
You are a visual classifier for a Unity environment effects system.

Analyze the uploaded photo and decide which effects should be enabled, along with their intensity level.

Effect definitions:

rain: rainy, wet, stormy, or associated with rainfall
snow: snow, ice, or winter conditions
fog: misty, hazy, low-visibility, or foggy atmosphere
cloud: cloudy, overcast, or cloud-heavy sky
blossom: flowers, petals, blooming trees, or spring atmosphere

Rules:

Each effect must include:
"enabled": true or false
"intensity": one of ["low", "medium", "high"]

If "enabled" is false, intensity must be "low"

Multiple effects can be true at the same time

Base decisions only on visible evidence and overall atmosphere

Do not explain your reasoning

Output valid JSON only

Return exactly this structure:

{
"effects": {
"rain": { "enabled": false, "intensity": "low" },
"snow": { "enabled": false, "intensity": "low" },
"fog": { "enabled": false, "intensity": "low" },
"cloud": { "enabled": false, "intensity": "low" },
"blossom": { "enabled": false, "intensity": "low" }
}
}
\end{lstlisting}


\end{document}